\documentstyle[epsfig]{mn}

\title[AGB stars in Leo I]
{The Brightest AGB Stars in the Leo I Dwarf Spheroidal Galaxy}
\author[Menzies et al.]
{John Menzies$^{1}$, Michael Feast$^{2}$, Toshihiko Tanab\'{e}$^{3}$, 
Patricia Whitelock$^{1}$ \cr  
and Yoshikazu Nakada$^{4}$\\
$^{1}$ South African Astronomical Observatory, P.O.Box 9, Observatory,
7935, South Africa. jwm,paw@saao.ac.za\\
$^{2}$ Astronomy Department, University of Cape Town, Rondebosch, 7701,
South Africa. mwf@artemisia.ast.uct.ac.za\\
$^{3}$ Institute of Astronomy, School of Science, The University of Tokyo,
Mitaka, Tokyo 181-8588, Japan. ttanabe@ioa.s.u-tokyo.ac.jp\\
$^{4}$ Institute of Astronomy, School of Science, The University of Tokyo,
Mitaka, Tokyo, 181-8588, Japan. nakada@kiso.ioa.s.u-tokyo.ac.jp\\}

\begin{document}

\maketitle

\begin{abstract}
The first results of a study of the dwarf spheroidal galaxy, Leo I,
using the new Nagoya-South African Infrared Survey Facility (IRSF)
are presented.  $J,H,K_{s}$ observations show that 
most, if not all, of at least the top magnitude of the AGB in $K_{s}$
 is populated
by carbon stars.
In addition there are five very red objects which are believed to be
dust enshrouded AGB stars. One of these is, remarkably, well
outside the main body of the galaxy.
Three of these obscured stars and five known carbon stars show variability in
observations 11 months apart. One of the obscured stars has 
$\Delta K_{s} = 0.87$ making it highly likely that it, at least, is a
Mira variable.
The tip of the AGB is at $M_{bol} \sim -5.1$, but further variability
studies are necessary to obtain a definitive value.
Comparison with carbon stars, both Miras and non-Miras, in Magellanic Cloud 
clusters and
taking into account other evidence on the ages and metallicities of
Leo I populations suggests that these obscured stars belong to the youngest
significant population of Leo I and have ages of $\sim 2$ Gyr.

\end{abstract}

\begin{keywords}
galaxies: dwarf - galaxies: stellar content - stars: AGB and post-AGB -
stars: variable: other - Local Group
\end{keywords}

\section{Introduction}
A programme has been started, using the 
recently commissioned Nagoya-South African
1.4m Infrared Survey Facility (IRSF) at SAAO Sutherland, to study the stellar
populations, evolution and structures of Local Group galaxies. One aim of
this programme is to detect long period variables (Miras and other types)
in these systems
and to derive their infrared light curves. The programme will necessarily
take several years to complete. In the present communication we discuss
the light that initial observations of the dwarf spheroidal galaxy, Leo I,
throw on the AGB star population of that galaxy.

\section{Observations}
The IRSF
is a 1.4-m telescope constructed and operated in terms of an agreement
between SAAO and the Graduate School of Science and School of Science,
Nagoya University, to carry out
specialized surveys of the southern sky in the infrared.
The telescope is equipped with a 3-channel camera, SIRIUS,
constructed jointly by  Nagoya University 
and the National Astronomical Observatory of Japan  (Nagashima et al. 1999), 
that 
allows {\it J, H\ } and {\it K$_s$} images to be obtained simultaneously.
 The field of view is 7.8 arcmin square with a scale of 0.45
arcsec/pixel.

Images centred on Leo I (referred to hereafter as field A) 
were obtained at two epochs, 2001-01-16 and
2001-12-19, and processed
by means of the standard IRSF pipeline (Nakajima, private communication).
A single image comprises 10 dithered 30-s exposures.  Three such
sets of frames were combined to give an effective 900-s exposure in 
each of {\it J, H\ } and $ K_{s}$ at both epochs. 
At this stage, the effective field of view is reduced to 7.2 arcmin square.
 Standard stars from Persson et al. (1998) were observed on each night
and the results presented here are in the natural system of 
the SIRIUS camera, but with the zero point of the Persson et al. standards. 
At the first epoch, we obtained a supplementary set of images of an
adjacent field (field B) centred 7 arcmin to the east of Field A. The two
fields overlap by only about 20 arcsec.

Photometry was carried out on the images with the aid of DoPHOT (Schechter,
Mateo \& Saha 1993) used in fixed-position mode.
Since the seeing was much better at the first epoch
(1.6 arcsec as opposed to 2.6 arcsec at the second epoch), 
the $K$ image obtained then
was used as a template to measure a complete sample of stars to a
limiting magnitude of about $K_{s} = 16.0$.  
The data are plotted in Figs. 1 ($K_{s}$ vs $(J-K_{s})$) and 2 
($(J-H)$ vs $(H-K_{s})$).  

\begin{figure*}
\epsfig{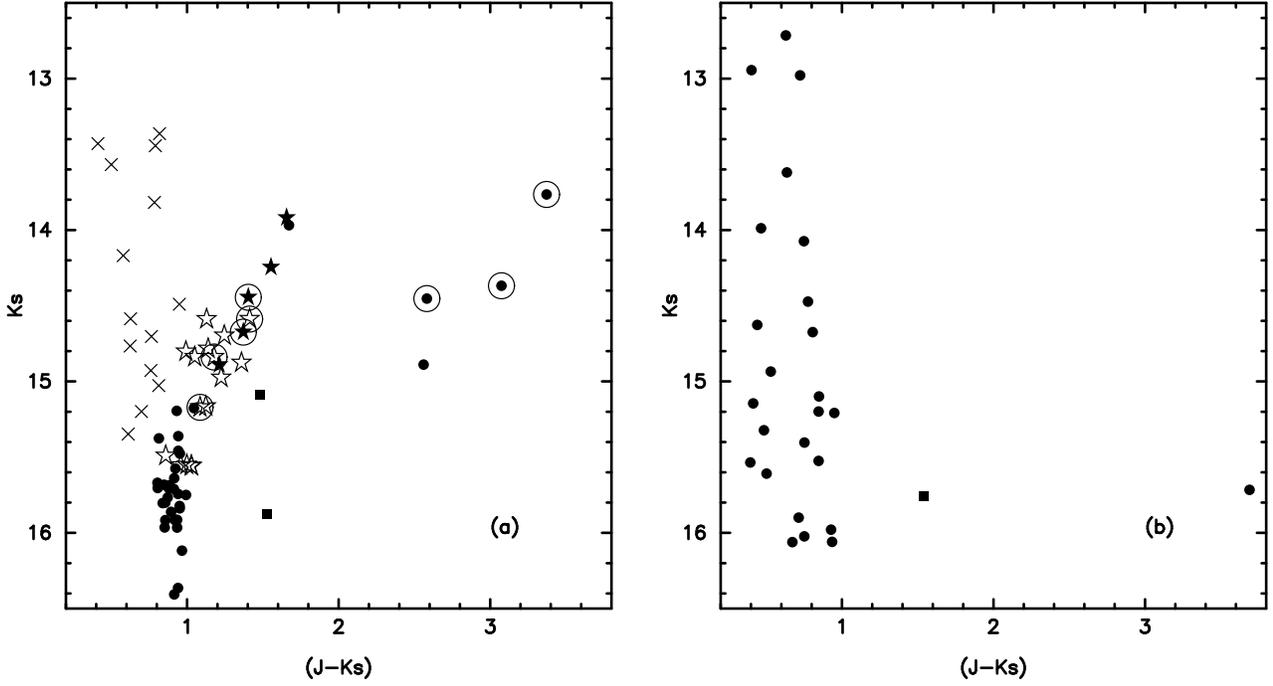}
\caption{ \it \small{(a) K$_s$ vs (J-K$_s$) for Leo I centre . Crosses are 
probable field stars,
open star symbols denote ALW carbon stars, while filled star symbols refer
to DB stars not measured by ALW. Filled squares represent probable
background galaxies. Variables from Table 2 are circled. (b) K$_s$ vs
(J-K$_s$) for a field 7 arcmin east of the centre of the galaxy. The
nearly vertical sequence is probably due to field stars, the square is
probably a field galaxy, and star E is at the lower right.
}}
\end{figure*}

\begin{figure*}
\epsfig{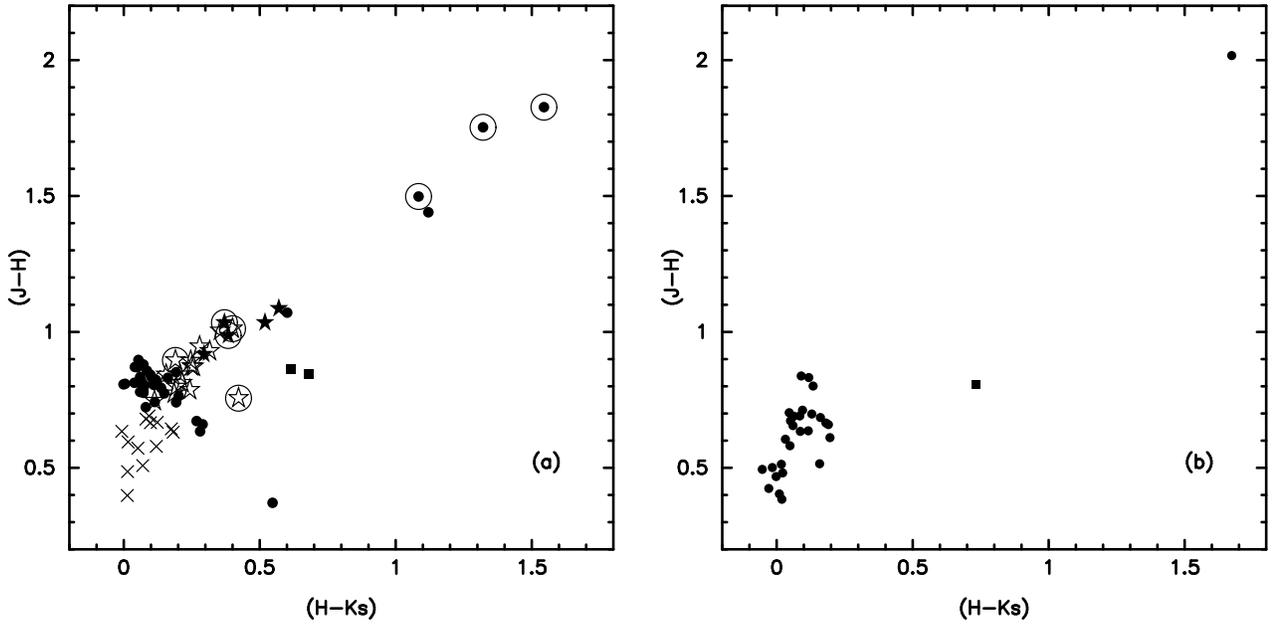}
\caption{\it \small{(J-H) vs (H-K$_s$) for (a) Leo I centre and (b) a field 7
arcmin east. Symbols as in figure 1.
}}

\end{figure*}

In the past, $E_{(B-V)} =0.02$, derived from Burstein and Heiles (1984)
has generally been adopted for this galaxy (e.g. Lee et al. 1993). The results
of Schlegel et al. (1998) suggest that a larger value ($\sim 0.04$) is
appropriate. In neither case will this lead to significant reddening at
JHK and we have neglected it.

\subsection{Probable Field Stars}
The stars lying to the blue of the main concentration of stars in Fig. 1(a)
are shown as crosses  there and are similarly marked in Fig. 2 (a).
They are likely to be foreground field stars.  This view is strengthened by
the results for the adjacent field B where the stars in the almost vertical 
sequence are almost certainly field dwarfs. 
Two points 
(filled squares) at $(J-K_{s}) \sim$1.5
in Fig. 1(a) and one in Fig. 1(b) are likely, from their 
colours, to be due to background galaxies. 
Indeed, close inspection of our
images shows evidence for extended emission associated with two of them,
one of which is clearly a galaxy on publicly available HST images. 

\subsection{The AGB sequence}
Apart from the field stars discussed above and the four very red objects 
discussed in the next section, all the stars in field A lie on a sequence in 
Fig. 1(a). Objects identified as carbon stars by Azzopardi, Lequeux \&  
Westerlund (1986 = ALW) or by Demers \& Battinelli (2002 = DB) are indicated
by star symbols.
Photometry was obtained for 21 known or suspected carbon stars in 
Leo I, which account for all the stars in the ALW and DB lists  except for 
the following: DB 4 and 8 which are seen on the edges of our frames but were 
not measured; DB 13 and ALW 4 and 6 which are outside our fields.\\ 

Using the bolometric corrections for carbon stars as a
function of $(J-K)$ given by Frogel, Persson and Cohen (1980) and a distance
modulus of 22.2 for Leo I based on the RGB tip (Lee et al. 1993)
\footnote{We have preferred to use this distance modulus rather than  that 
given recently
by Held et al. (2001) ((m-M) = 22.0) which depends on an adopted RR Lyrae
absolute magnitude.}
one finds that the carbon star sequence runs from $M_{bol} = -4.2$ 
at $(J-K_{s}) = 0.9$ to $M_{bol} = -5.1$ at $(J-K_{s}) = 1.7$. However,
as can be seen from work on galactic carbon stars (Whitelock 2000),
the stars at the redder end of this sequence may well be Mira variables
and cannot be taken as defining the upper limit of the sequence without
more observations. All the stars of this sequence are AGB stars. 
The RGB tip is expected to be fainter than $M_{bol} = -4.0$ for any
reasonable metallicities or ages (see for instance Castellani et al. 1992,
Salaris \& Cassisi 1998).  The present results show clearly
how the blue-green grism results of ALW miss the brighter carbon stars and 
would therefore lead to an underestimate of the brightness of the AGB tip. A
similar underestimate of the AGB tip is present in {\it VI} work (e.g.
Lee et al. 1993, fig 4d).
All but one of the brightest, reddest objects constituting the top of the
AGB sequence appear in the DB list, and it is interesting to note that the 
obscured objects discussed below would, when dereddened, extend this sequence 
to even brighter K$_s$ magnitudes.\\

At the lower (bluer) end of the AGB sequence in Fig. 1(a) (which is of course
determined by our adopted magnitude cut off) there is a group of objects without
spectral classification. They lie mainly to the blue of the known carbon stars
in Fig 2(a). It would be interesting to know whether these are O or C rich
objects. A few of them may be foreground stars.\\

Fig 1(a) contains an object, 
without spectral classification,
near the top of the AGB sequence with
$(J-K_{s}) = 1.66$.
In view of its position in Figs 1(a) and 2(a) it seems likely that it is
also a carbon star. The fact that it was not found in the survey of DB
may mean that it is a variable and was  below the magnitude limit of
DB at the time of their observations. The star's colour and luminosity  are
similar to those expected for carbon Miras.

\subsection{Obscured Objects}
Four very red objects are conspicuous in Figs 1(a) and 2(a) (field A) while
there is an even redder one, though it is rather fainter in K$_s$, in field
B (Figs 1(b) and 2(b)).  The positions and photometry of these five
objects at JD 2451962.52 are listed in Table 1.
Their locations in the colour-magnitude and two-colour diagrams
are consistent with their being bright AGB stars obscured by
circumstellar dust (see e.g. Whitelock 2000). 
In view of the
fact that the top magnitude or more of the AGB sequence just discussed is 
heavily, and perhaps entirely, populated by carbon stars, it seems very likely
that these obscured stars are  also carbon stars. 
As noted by Nikolaev \& Weinberg (2000) several other types of object have
colours that would put them in this part of the two-colour diagram, for
example, OH/IR stars and protostars. For the reasons outlined above, it
seems more likely they are carbon- rather than oxygen stars, though the
discussion on ages below is not significantly affected if the latter is the
case. The presence of protostars of this brightness in a dwarf spheroidal
galaxy would be quite remarkable. 

These five stars
are the most extreme examples yet found of this type of obscured object
in dwarf spheroidal galaxies. 
The nearest comparable objects, though they are somewhat bluer($(J-K) \sim
2.4$), are the two carbon Miras that have been found in
the Sagittarius dwarf spheroidal (Whitelock et al. 1999).

\begin{table}
\caption{Photometry and positions 
(equinox 2000)
of  5 very red objects in the field of Leo I}
\begin{tabular}{cllllll}
name & \multicolumn{1}{c}{R.A.} & \multicolumn{1}{c}{Dec} &
\multicolumn{1}{c}{K$_s$} & \multicolumn{1}{c}{J-H} &
\multicolumn{1}{c}{H-K$_s$}
& \multicolumn{1}{c}{J-K$_s$}
\\
A &10:08:29.3 &12:18:52 &14.37 &1.75 &1.32 &3.07 \\
B &10:08:27.3 &12:18:57 &14.45 &1.50  &1.08 &2.58 \\
C &10:08:22.3 &12:17:57 &13.76 &1.83 &1.54 &3.37 \\
D &10:08:41.3 &12:18:08 &14.89 &1.44 &1.12 &2.56 \\
E &10:09:00.5 &12:19:01 &15.72 &2.02 &1.67 &3.69 \\

\end{tabular}
\end{table}

\subsection{Variability}

A comparison of the data from the two epochs of observation (JD 2451926.52
and 2452263.55) shows that a number of the brighter  stars are variable. 
We list these stars in Table 2, where we give the magnitude difference
between the second and first epochs, together with the standard deviation
of the difference based on the internal errors of measurement as given by
the DoPHOT program. Stars where variations of at least 4$\sigma$ occur in 
two or more of the passbands are considered to be variables.
Three of the obscured stars are included, together with five carbon stars:
two from the ALW list, two in common between ALW and DB,
and one in the DB list only.
There are hints that some of the other carbon stars are also
variable. Note that 
the obscured
star C must have a very large amplitude and probably a
period of at least 1 year. This is consistent with the fact that stars 
with JHK colours similar to these stars are generally Mira variables
(see Whitelock 2000, fig. 1).

\begin{table}
\caption{Probable variable stars}
\begin{tabular}{crrrrrr}
name & $\Delta J$ & $\sigma_J$ &$\Delta H$ & $\sigma_H$ &$\Delta K$ &
$\sigma_K$ \\
C     & 0.92  & 0.16 & 0.96  & 0.06  & 0.87  & 0.03  \\
ALW9  & 0.15  & 0.03 & 0.08  & 0.03  & 0.20  & 0.04     \\
B     & 0.34  & 0.09 & 0.18  & 0.04  & 0.14  & 0.03    \\
C10   & 0.22  & 0.03 & 0.15  & 0.03  & 0.14  & 0.04    \\
C06   & -0.19 & 0.02 & -0.14 & 0.02  & -0.14 & 0.03  \\
ALW16 & -0.24 & 0.03 & -0.02 & 0.03  & -0.18 & 0.04  \\
C07   & -0.28 & 0.02 & -0.16 & 0.02  & -0.24 & 0.02  \\
A     & -0.44 & 0.07 & -0.26 & 0.03  & -0.26 & 0.02  \\
\end{tabular}
\raggedright
Obscured stars are denoted by their letters in Table 1, other stars
by their DB numbers (C10 etc.) or their ALW designation.
\end{table}

\section{Discussion}
 A full discussion of the nature of the obscured  stars must obviously be 
deferred till further variability studies have been made. 
Neverthless it is of interest to 
draw some preliminary conclusions.\\

The results we have obtained are strikingly similar to those of
Nishida et al. (2000) (see also Tanab\'{e} et al. 1997, 1999) 
who found an obscured carbon Mira in  each of the intermediate age clusters,
NGC 419, NGC 1783 and NGC 1978,
in the Magellanic Clouds.
These variables have values of $(J-K)$ (SAAO system)
between 3.75 and 4.76. Thus, in colour, the obscured Leo I stars lie
between the Sagittarius dwarf spheroidal Miras,
mentioned above, and the LMC cluster Miras. 
In the Magellanic Cloud clusters the tip of the unobscured AGB is
at $M_{K} \sim -8.4$ and $M_{bol} \sim -5.3$ (using data from
Frogel et al. 1990 and distance moduli of 18.6 and 19.0 for the
LMC and SMC). In Leo I the corresponding values are --8.3 and --5.1
which however remain uncertain pending full variability studies.\\

In the Magellanic Cloud clusters the Miras have a mean $M_{K}$ of --8.0 and
$M_{bol} = -5.1$ or slightly brighter.
The corresponding figures for the obscured stars in Leo I are
quite uncertain both because three, and possibly all, are variable and
also because the estimation of bolometric corrections for such stars from 
$JHK$ colours is rather uncertain. The three obscured stars found
to be variable
have $M_{K} = -8.0$ and $M_{bol} \sim -4.9$ whilst all five obscured
stars yield $M_{K} = -7.6$ and $M_{bol} \sim -4.6$. To derive these
results for Leo I  we have used a relation of $(H-K)$ to bolometric
correction for carbon Miras derived by Whitelock (to be published) which
includes the use of ISO data. Thus the tip of
the unobscured AGB in $M_{K}$ and $M_{bol}$ is very similar in Leo I
to that in the three Magellanic Cloud clusters. The obscured AGB may be 
fainter in $M_{bol}$ than the cluster Miras but that is not certain.\\

It is of some interest to note that if the intrinsic (underlying) colours
of the five obscured stars were $(J-K_{s}) \sim 2.0$, they  would all move
to a position near the top of the AGB in $K_{s}$, 
when dereddened using
the reddening law ($\Delta K \sim \Delta (J-K)$) found for the
circumstellar envelope of the
galactic carbon Mira R For (Feast et al. 1984).\\

The Magellanic Cloud clusters are estimated to have ages in the
range 1.6 to 2.0 Gyr, metallicities, [Fe/H] $\sim -0.6$, and turn-off
masses of $\sim 1.5 M_{\odot}$. 
In view of the above discussion it seems likely that the Leo I obscured
stars are in the same age and mass range.
It is therefore interesting to note
that Gallart et al (1999) suggest that major star formation in Leo I stopped
about 2 Gyr ago and that a metallicity of that population 
as high as [Fe/H] $\sim -0.6$
is possible. The results thus suggest that the obscured stars, and 
the most luminous unobscured AGB stars, at least,
belong to this youngest major stellar component of Leo I.\\

The very red star (star E) in field B is especially interesting. It is
$\sim 8$ arcmin from the centre of Leo I, much further out than any of the
other stars we discuss. Even the C stars outside field A, mentioned in
section 2.2, are within $\sim 4.5$ arcmin of the centre. There is HI in
a wide area around Leo I and apparently associated with it (Blitz \&
Robishaw 2000). Star counts (Irwin \& Hatzidimitriou 1995) show that
Leo I has a tidal radius of 13 arcmin, which extends well beyond the 
distance of star E from the centre. However
the stellar density at these distances is very low compared with that in
the main body of the galaxy. Furthermore, in dwarf spheroidals the
younger populations are generally more concentrated to the centre
than the older ones (e.g. Harbeck et al. 2001). Thus if we have
correctly interpreted this star as similar to the obscured AGB stars in
field A, and hence relatively young, its position at such a large distance
from the centre is remarkable. This is particularly so when it is recalled
that such objects are short lived and are therefore tracers of much
larger populations. Since the star has only been observed at one epoch
we cannot comment on its possible variability\footnote{Since this paper
was submitted, further images have been obtained which show that star E is
variable, which strengthens the assumption that it is an obscured AGB
star.}.

\section{Conclusions}
We have obtained JHK$_s$ photometry of a complete sample of stars to
K$_s$=16 in a 7.2 arcmin square field centred on Leo I and in an 
adjacent field. This sample includes
all 21 known carbon stars falling in the imaged area.
  Our results show that the top one magnitude or more of the AGB in $K_{s}$
is populated entirely or almost entirely by carbon stars. 
These stars form a sequence in the
$K_{s} - (J-K_{s})$ diagram and several of them are variable.
In addition there are five very red
stars, at least three of them variable, 
which from their magnitudes and colours are deduced to be 
AGB tip stars obscured by dust shells. They are strong candidates
for Mira variability. These stars, 
and at least the brightest unobscured AGB stars,
probably belong to the youngest
and most metal rich of the significant stellar populations in Leo I.
Comparison with carbon stars in Magellanic Cloud clusters suggests
ages of about 2 Gyr for these stars in agreement with the age of the
youngest major population in this galaxy as derived in other ways.
Surprisingly, in view of the fact that younger populations are generally
more centrally concentrated than others in dwarf spheroidals, one
of the obscured stars lies about 8 arcmin from the centre of the galaxy,
compared with a tidal radius of 13 arcmin.
\subsection*{Acknowledgements}
  We are grateful to Noriyuki Matsunaga for help at the telescope.
We also thank Yasushi Nakajima for providing the reduction pipeline
and for information on the photometric system. The IRSF project was
initiated and supported by Nagoya University and the National Astronomical
Observatory of Japan, and we thank Professor Sato, Professor Nagata 
and all others involved in the project .

\end{document}